\documentclass[pra,twocolumn,superscriptaddress,amsmath,amssymb]{revtex4-1}

\usepackage{bm}

\usepackage{graphicx}

\begin{document}

\title{Exceeding the Landau Speed Limit with Topological Bogoliubov Fermi Surfaces}

\author{S. Autti}
\affiliation{Department of Applied Physics, Aalto University, POB 15100, FI-00076 AALTO, Finland}\affiliation{Department of Physics, Lancaster University, Lancaster LA1 4YB, UK}

\author{J.T. M\"{a}kinen}
\affiliation{Department of Applied Physics, Aalto University, POB 15100, FI-00076 AALTO, Finland}
\affiliation{Department of Physics, Yale University, New Haven, Connecticut, 06520, USA}
\affiliation{Yale Quantum Institute, Yale University, New Haven, Connecticut, 06520, USA}

\author{J. Rysti}
\affiliation{Department of Applied Physics, Aalto University, POB 15100, FI-00076 AALTO, Finland}

\author{G.E. Volovik}
\affiliation{Department of Applied Physics, Aalto University, POB 15100, FI-00076 AALTO, Finland}\affiliation{L.D. Landau Institute for Theoretical Physics, Moscow, Russia}

\author{V.V. Zavjalov}
\affiliation{Department of Applied Physics, Aalto University, POB 15100, FI-00076 AALTO, Finland}\affiliation{Department of Physics, Lancaster University, Lancaster LA1 4YB, UK}

\author{V.B. Eltsov}\email{vladimir.eltsov@aalto.fi}
\affiliation{Department of Applied Physics, Aalto University, POB 15100, FI-00076 AALTO, Finland}


\begin{abstract}
A common property of topological systems is the appearance of topologically protected zero-energy excitations. In a superconductor or superfluid such states set the critical velocity of dissipationless flow $v_{\mathrm{cL}}$, proposed by Landau, to zero. We check experimentally whether stable superflow is nevertheless possible in the polar phase of p-wave superfluid $^3$He, which features a Dirac node line in the energy spectrum of Bogoliubov quasiparticles. The fluid is driven by rotation of the whole cryostat, and superflow breakdown is seen as the appearance of single- or half-quantum vortices. Vortices are detected using the relaxation rate of a Bose-Einstein condensate of magnons, created within the fluid. The superflow in the polar phase is found to be stable up to a finite critical velocity $v_{\rm c}\approx 0.2\,$cm/s, despite the zero value of the Landau critical velocity. We suggest that the stability of the superflow above $v_{\mathrm{cL}}$ but below $v_{\rm c}$ is provided by the accumulation of the flow-induced quasiparticles into pockets in the momentum space, bounded by Bogoliubov Fermi surfaces. In the polar phase this surface has non-trivial topology which includes two pseudo-Weyl points. Vortices forming above the critical velocity are strongly pinned in the confining matrix, used to stabilize the polar phase, and hence stable macroscopic superflow can be maintained even when the external drive is brought to zero.
\end{abstract}

\maketitle

\def\vec#1{\mathbf{#1}}
\def\unit#1{\hat{\vec{#1}}}
\def\rhon{\rho_{\rm n}}
\def\rhos{\rho_{\rm s}}
\def\vc{v_{\rm c}}
\def\vvs{\vec v_{\rm s}}
\def\vs{v_{\rm s}}

\section{Introduction}

The stability of superflow in superfluids and superconductors is supported by both topology and the Landau criterion. Via quantization of circulation, the topological stability protects gradual decay of flow around vortices and in a ring geometry. The Landau criterion protects the superflow against decay via creation of quasiparticles for velocities below the Landau critical velocity $v_{\rm cL} = \min [E(\vec p)/p]$, where $E(\vec p)$ is the energy spectrum of quasiparticles with momentum $\vec p$. In Fermi superfluids and superconductors $v_{\rm cL} \approx \Delta /p_{\rm F}$, where $\Delta$ is the gap in the energy spectrum of Bogoliubov quasiparticles and $p_{\rm F}$ is the Fermi momentum.  In topological systems, appearance of sub-gap (in particular, zero-energy) states or presence of nodes in the energy gap is ubiquitous. How such zero-energy states affect the stability of superflow in topological superconductors and superfluids is an open question.

Remarkably, in the topological superfluid phases of $^3$He, superflow may persist when one \cite{Bradley2016,Thuneberg2018}, or even both \cite{Ruutu1997,AndersonToulouse1977,Tsvelik2017}, of those constraints are violated. In particular, topological protection is absent in the chiral superfluid $^3$He-A, where the circulation is topologically unstable towards a phase slip with the formation of skyrmions \cite{AndersonToulouse1977,Tsvelik2017}. However, contrary to the statement in Ref. \cite{Tsvelik2017}, the superflow persists up to velocity $v\sim 0.1$cm/sec \cite{Ruutu1997}. This velocity also exceeds the Landau critical velocity, which is zero in  $^3$He-A due to presence of two point nodes $[E(\vec p)=0]$ in the energy spectrum. On one hand, the absence of topological stability does not exclude local stability of superflow, supported by anisotropy of the superfluid density, effects of boundaries, applied magnetic field, or spin-orbit interaction. On the other hand, superflows exceeding the Landau critical velocity do not necessarily lead to the destruction of superfluidity in Fermi superfluids \cite{Volovik2003}. In the super-Landau superflow some Bogoliubov quasiparticle states acquire negative energy. Fermionic quasiparticles start to occupy those energy levels forming a Fermi surface. Such a Fermi surface is called the Bogoliubov Fermi surface (BFS). In superfluid $^3$He and in cuprate superconductors \cite{Volovik93,Moler94,Wang2001,Stewart2017} the BFS appears in the presence of superflow, while in  systems with multiband energy spectrum or with broken time-reversal symmetry the BFS may exist even in the absence of superflow \cite{Volovik1989,Liu2003,Agterberg2017,LiangFu2018,Brydon2018,Sumita2019,Zyuzin2019,Setty2020}. Note that in a nodal topological superfluid or in a cuprate superconductor the superflow explicitly breaks time-reversal and inversion symmetries, and thus origin of the BFS can be considered on a common ground in different systems.

Appearance of the BFS gives rise to a non-zero density of states at zero energy and thus to a non-zero normal component density $\rhon$ even at $T=0$. When all the negative states are occupied, the equilibrium value of $\rhon(T=0)$ is reached, and the non-dissipative superflow is restored, though with smaller superfluid density, $\rhos(T=0)=\rho-\rhon(T=0)$. The superflow above the Landau critical velocity remains stable until some other critical velocity $\vc$ is reached. This can be either the velocity at which $\rhon(T=0)=\rho$ and thus the superfluid density $\rhos=\rho-\rhon$ vanishes, or the critical velocity at which quantized vortices or other topological defects, such as skyrmions, are created. 

\begin{figure}[t]
\centerline{\includegraphics[width=1.0\linewidth]{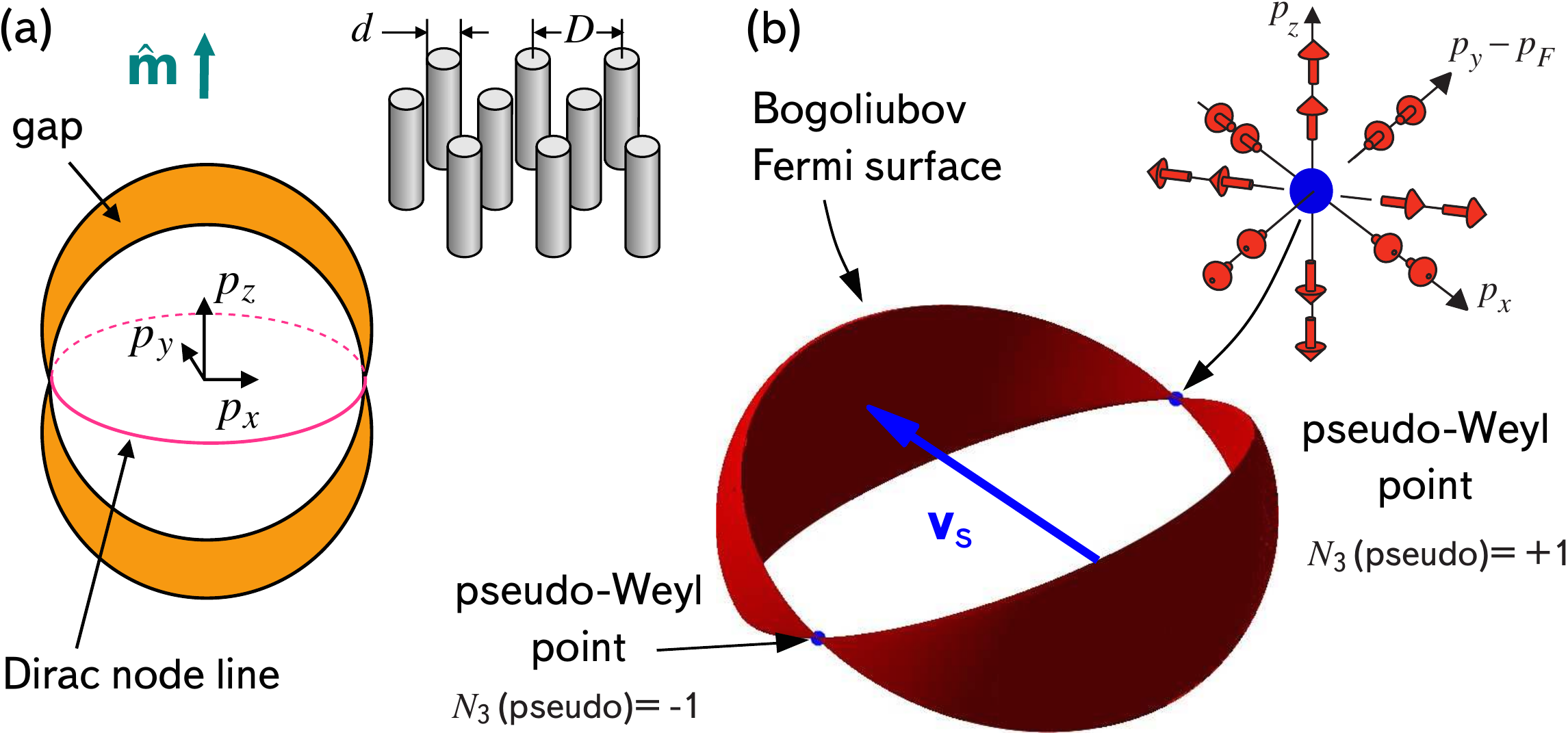}}
\caption{The polar phase of superfluid $^3$He engineered with nanostructured confinement. (a) The confining matrix is a set of parallel solid strands, realized using commercial nafen material with $d\approx9\,$nm and $D\approx 35\,$nm \cite{PolarDmitriev}.  In the stationary polar phase, the energy spectrum of Bogoliubov quasiparticles includes a Dirac node line in the plane perpendicular to strands. (b)~In the presence of superflow $\vvs$ the node line transforms to the Bogoliubov Fermi surface, consisting of
two Fermi pockets, which touch each other. Here  the superflow is applied along the $x$ axis, and touching points at $\vec p =\pm p_F \unit y$
are pseudo-Weyl points. Their topology is illustrated as the hedgehog in momentum space, with the topological invariant in Eq.~(\ref{InvariantPseudo}). Arrows show direction of the $\unit n$ vector and
the parameters in Eq.~(\ref{epsilon}) are chosen as $m^*c/p_F = 1/12$ and $\vs/c =1/2$.
\label{polarfig}}
\end{figure}

The topology and other properties of the p-wave superfluid $^3$He can be tuned on a wide range via controlling temperature, pressure, or magnetic field \cite{VollhardtWolfle1990}, or by introducing engineered nanoscale confinement \cite{LevitinScience,HalperinNatPhys,Levitin2019,Shook2020}.  Recently a new phase of $^3$He, the time reversal symmetric polar phase has been engineered using such confinement \cite{Aoyama2006,PolarDmitriev,Dmitriev2018}. The polar phase, Fig.~\ref{polarfig}, features a Dirac nodal line, robust to disorder and impurities owing to the extension of the Anderson theorem \cite{FominAnderson,PolarT3,IkedaAnderson}. Due to the presence of the nodal line with $E(\vec p)=0$, the Landau criterion in the polar phase is violated for any non-zero velocity.

The purpose of the current Report is twofold: First, we experimentally demonstrate that the superflow in the presence of the nodal line remains stable until the fluid velocity at the sample boundaries reaches $0.24\,$cm/s, well above the zero Landau critical velocity. At higher velocity the flow, driven by rotation of the sample container, becomes unstable towards formation of quantized vortices. The appearance of vortices, strongly pinned to the strands of the confining matrix, is detected as the increased relaxation rate of a Bose-Einstein condensate of magnon quasiparticles \cite{MagnonBEC,BECpolar}. Vortices remain in the sample for days after the rotation is stopped maintaining long-living superflow exceeding the Landau critical velocity even in a stationary sample. These observed features of vortex dynamics in the polar phase are supported by numerical simulations. Second, we discuss the topology of the resulting Bogoliubov Fermi surface and provide suggestions for characterization of the effects of the BFS on superfluid properties in future experiments.

\section{Polar phase}

The order parameter in the polar phase is
\begin{equation}
A_{\mu j}=\Delta_P{\bf \hat d}_\mu {\bf \hat m}_j e^{i\phi}\,.
\end{equation}
Here, $\Delta_P$ is the maximum gap in the quasiparticle energy spectrum, $\phi$ is the superfluid phase, and ${\bf \hat d}$ is a unit vector of spontaneous anisotropy in the spin space. Orbital anisotropy vector ${\bf \hat m}$ is locked along the nafen strands, while the node line in the energy spectrum develops in the plane perpendicular to the strands, Fig.~\ref{polarfig}a. In our sample the strands are oriented along the rotation axis, labeled $\hat{z}$. 

\begin{figure}[t]
\centerline{\includegraphics[width=0.9\linewidth]{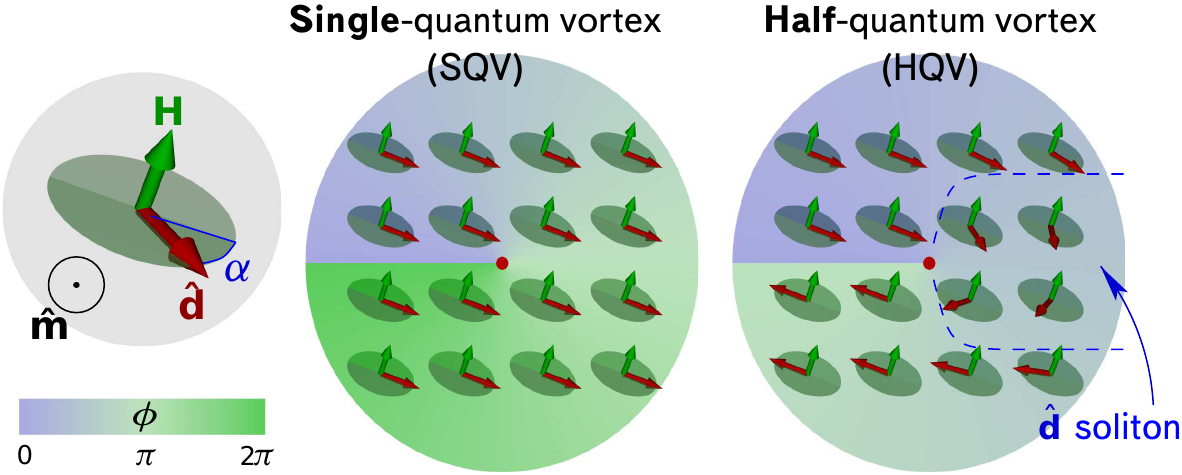}}
\caption{Types of quantized vortices in the polar phase. The order parameter phase $\phi$ (background color) winds by $2\pi$ around a single-quantum vortex, and by $\pi$ around a half-quantum vortex. To keep the order parameter single-valued, vector $\unit d$ (red arrows) also rotates around the HQV core, so that $\unit d \to -\unit d$ when $\phi \to \phi+\pi$. In non-axial magnetic field (green arrows) this leads to the formation of $\unit d$ solitons connecting HQVs pairwise (blue dash line). Vortex cores, vector $\unit m$, nafen strands, and the axis of rotation are perpendicular to the plane of the picture. \label{vorttypes}
} 
\end{figure}

This order parameter allows for both usual single-quantum phase vortices (SQV), around which $\phi \rightarrow \phi +2 \pi$, and half-quantum vortices (HQV), where $\phi \rightarrow \phi + \pi$ and $\alpha \rightarrow \alpha + \pi$~\cite{HQVs_prl}, Fig.~\ref{vorttypes}. Here $\alpha$ is the azimuthal angle of $\bf{ \hat d}$ in the plane perpendicular to the magnetic field. The $\unit d$ vector is kept in this plane by the Zeeman energy in the magnetic field of applied in our experiments. Additionally pure spin vortices with winding $\alpha \rightarrow \alpha + 2\pi$ and $\phi={\rm const}$ around the core can exist, but they are not relevant for the critical velocity in applied mass flow, so we do not discuss them here.

The quasiparticle energy spectrum in the polar phase, which is Doppler shifted in the presence of a superflow $\vvs$, takes the form
\begin{equation}
 E({\bf p})= \epsilon({\bf p}) + {\bf p}\cdot{\bf v}_{\rm s}\,\,, \,\,\epsilon^2({\bf p}) = c^2 p_z^2 + v_{\rm F}^2(p -p_{\rm F})^2
\,,
 \label{epsilon}
\end{equation}
where $c=\Delta_P/p_{\rm F}$. For any $\vvs$ not collinear with the $\hat z$ axis, this spectrum contains states with $E({\bf p})<0$ and the Landau critical velocity is zero, $v_{\rm cL} = 0$.

\begin{figure}[t]
\centerline{\includegraphics[width=\linewidth]{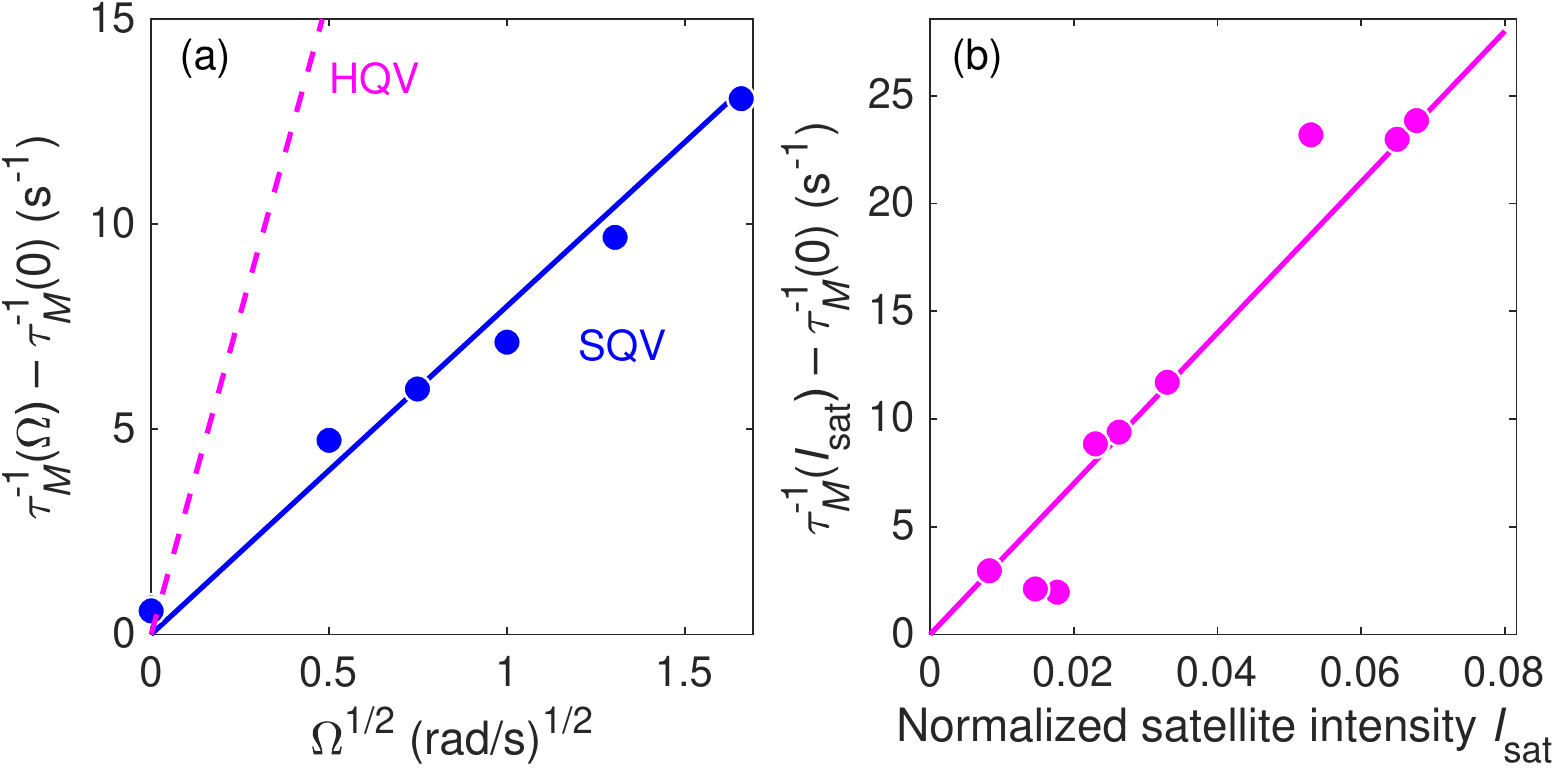}}
\caption{Increase of the relaxation rate of the magnon BEC $\tau_M^{-1}$ due to vortices in the polar phase. (a) The relaxation rate grows as a function of rotation velocity $\Omega$ applied at the transition to the superfluid state. For HQVs the shown slope is extracted from the data in Ref.~\cite{HQVs_prl} with the contribution of vortices, created by the Kibble-Zurek mechanism (KZM), removed. For SQVs KZM is suppressed by the symmetry-violating bias \cite{KZ_bias}. The measured points are shown by symbols and the line is a fit to $\Omega^{1/2}$ dependence. (b) For HQVs $\tau_M^{-1}$ (circles) is proportional to the total volume of the $\unit d$-solitons between HQV cores, measured by the area of the satellite peak $I_{\rm sat}$ in the normalized NMR spectrum, like in Fig.~\ref{spectra}. The line is a linear fit. \label{SQV_HQV_equilibrium}
} 
\end{figure}

\section{Measurements}

We study the stability of superflow starting from the initial state prepared by slowly cooling the stationary sample through the superfluid transition in a transverse magnetic field to suppress the formation of both HQVs and SQVs \cite{KZ_bias}. All measurements are performed on the $4\times4\times4$~mm$^3$ cubic sample container at 7 bar pressure and $T=0.4\,T_{\rm c}$ in the magnetic field $H=12\,$mT. The container is filled with nanomaterial called nafen, which consists of parallel columnar Al$_2$O$_3$ strands with 0.243\,g/cm$^3$ volume density. Then, at constant $T$, we gradually increase the rotation velocity in small steps, reducing the applied field to zero before changing the velocity.

When the change is finished, we restore the transverse magnetic field and measure the relaxation rate of long-living magnons, pumped to the sample by a radio-frequency excitation pulse. Under conditions of this work, magnons form a Bose-Einstein condensate (BEC) within the sample \cite{MagnonBEC,HPD,BECpolar}. The magnon BEC is manifested by coherent precession of magnetization with the same frequency and coherent phase, despite the inhomogeneity of the magnetic field or variation in the spin-orbit interaction strength. The precession slowly decays due to magnon loss. In $^3$He-B such condensates are thoroughly explored and were used as sensitive probes of temperature \cite{Lancrel,BECtemp}, collective modes \cite{LHiggs}, spin supercurrents \cite{HPDsscur}, analogue event horizon \cite{Skyba2019}, and of vortices \cite{HPDvortold,Hosio2013}. Their usefulness as sensors in the polar phase has not been known before this work.

We have found that the decay rate $\tau_M^{-1}$ of the magnon BEC is sensitive to the presence of vortices also in the polar phase, Fig.~\ref{SQV_HQV_equilibrium}. To calibrate this effect, we create an equilibrium array of vortices by rotating the sample at the angular velocity $\Omega$ while slowly cooling it down to the superfluid state. HQVs are produced with cooling in the zero or axial field \cite{HQVs_prl}. 

\begin{figure}[t]
\centerline{\includegraphics[width=\linewidth]{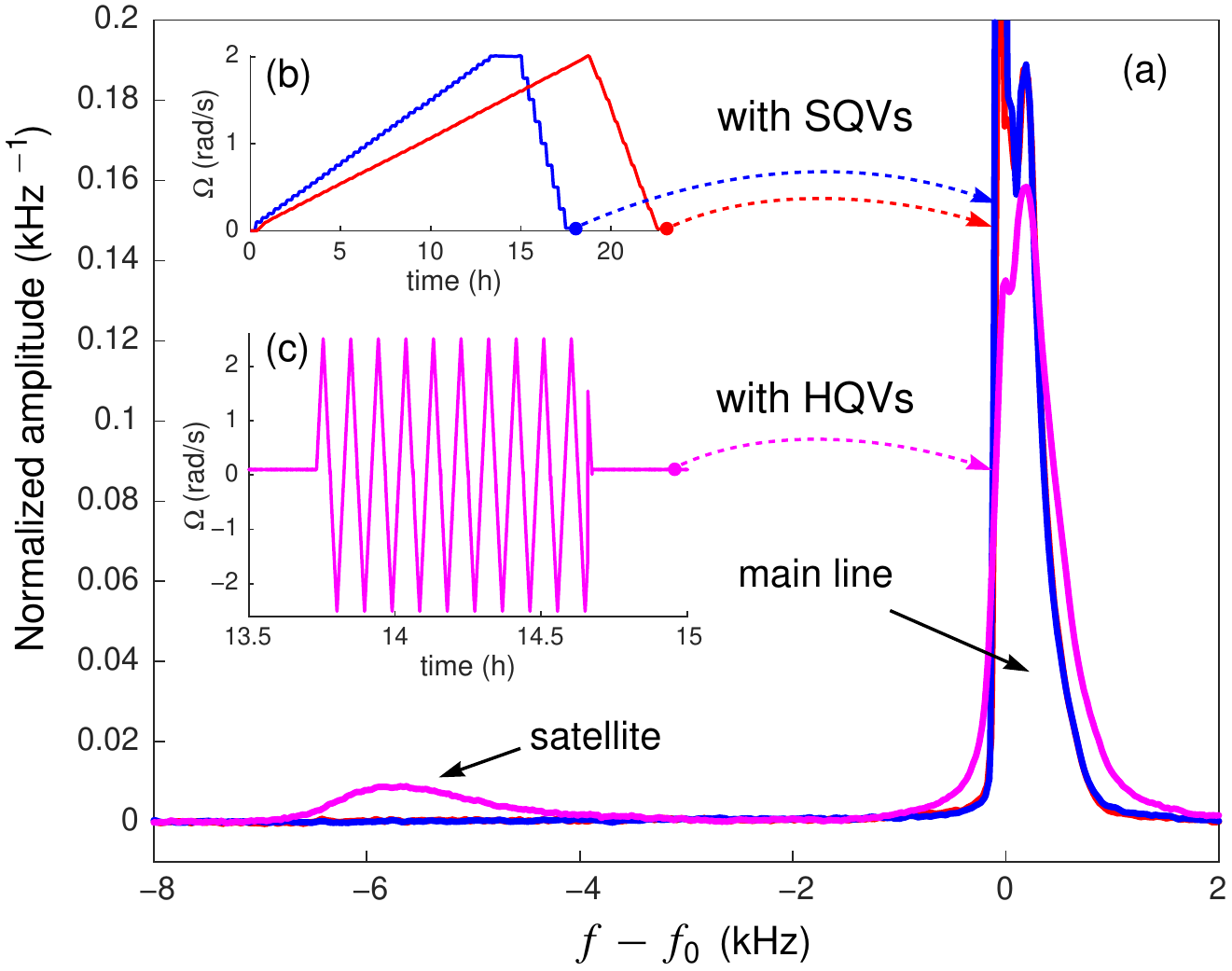}}
\caption{Continuous-wave NMR spectra of the polar phase measured in the magnetic field transverse to nafen strands with HQVs and SQVs present in the sample. (a) The absorption normalized to the total spectrum area is plotted versus the frequency shift from the Larmor value $f_0 = |\gamma|H/2\pi$, where $\gamma$ is the gyromagnetic ratio of $^3$He. The HQVs produce a satellite in the NMR spectrum, while for SQVs no clear distinguishing features are seen. (b) SQVs are produced with the slow sweep of the angular velocity $\Omega(t)$.  (c) Rapid changes in rotation velocity produce HQVs in addition to SQVs. Last of the ten periods of such drive applied in the course of the measurement in zero magnetic field is shown.}
\label{spectra}
\end{figure}

For HQVs, Fig.~\ref{SQV_HQV_equilibrium}b, $\tau_M^{-1}$ is proportional to the intensity $I_\mathrm{sat}$ of the characteristic satellite peak in the continuous-wave (cw) nuclear magnetic resonance (NMR) spectrum measured in the transverse field, Fig.~\ref{spectra}a. The $I_\mathrm{sat} \propto \Omega^{1/2}$ is essentially a fractional volume occupied by the $\unit d$ solitons connecting HQVs pairwise  \cite{HQVs_prl}. We conclude that the magnon BEC relaxation is concentrated in these solitons, where the spin configuration deviates from the equilibrium. 

If the magnetic field is applied transverse to the rotation axis and the nafen strands, HQVs become energetically unfavorable \cite{IkedaHQV}, and an array of SQVs is created on cooling through $T_{\rm c}$. SQVs cannot be easily identified based on cw NMR spectra, since SQVs are not associated with a soliton structure and their cores are too small to provide noticable contribution to the signal. However, a definite increase of the magnon BEC relaxation as a function of $\Omega$ is observed, Fig.~\ref{SQV_HQV_equilibrium}a. 

The origin of the magnon BEC relaxation in the B phase is conversion of the magnons from the condensate to longitudinal spin waves (light Higgs quasiparticles \cite{LHiggs}) in the distorted orbital texture surrounding non-axisymmetric vortex cores \cite{Laine}. In that case $\tau_M^{-1} \propto \Omega$ in the equilibrium vortex state. In the polar phase the structure of vortex cores is not known and expected to be substantially affected by the nafen strands with diameter $\sim 0.1$ of the vortex core diameter and spaced by a few times larger distance. Simultaneously, the orbital texture is believed to be pinned by the strands. Thus, the microscopic dissipation mechanism for coherent precession acting in the polar phase and the origin of the dependence $\tau_M^{-1} \propto \Omega^{1/2}$ remain unclear, but we nevertheless can use $\tau_M^{-1}$ as a marker for appearance of vortices. Moreover, usage of the magnon BEC probe proved to be essential for fulfilling the main goal of this work, establishing stability of the flow in the polar phase, as it turns out that the stability is lost with formation of SQVs (see the next section), which are invisible to classic linear NMR.

\section{Critical velocity}

\begin{figure}[t]
\centerline{\includegraphics[width=0.9\linewidth]{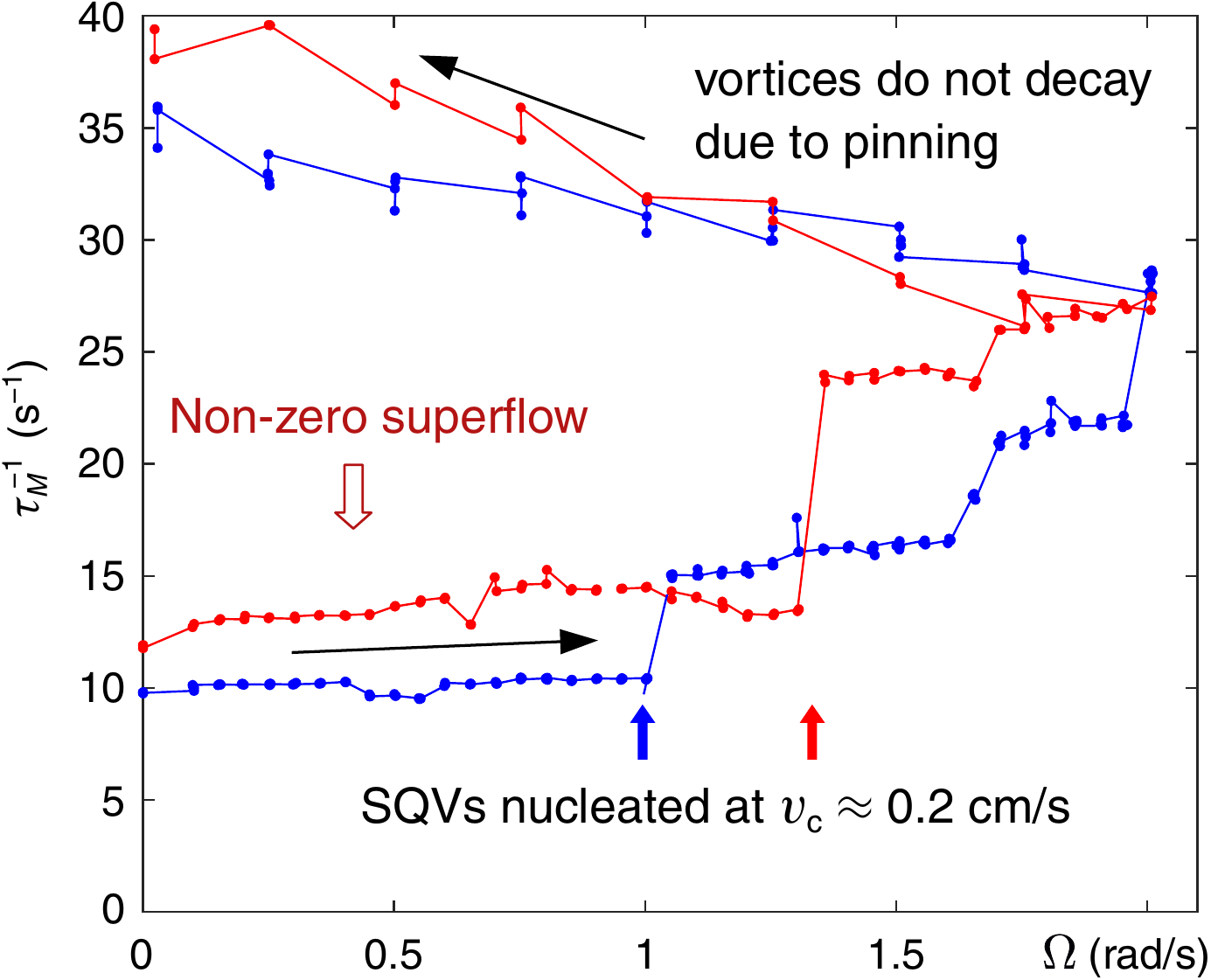}}
\caption{Change of the magnon BEC relaxation $\tau_M^{-1}$ when the rotation velocity $\Omega$ is gradually increased from 0 to 2\,rad/s and then decreased back to 0 at constant temperature, starting from the state with no vortices. Two traces for two independently prepared initial states are shown, the respective $\Omega(t)$ dependences and final spectra are plotted in Fig.~\ref{spectra}(b,a). Single-quantum vortices form first at $\Omega\approx 1\,$rad/s. Pinning prevents vortices from disappearance when $\Omega$ is decreased. Before vortex formation, stable superflow with velocity up to $v_{\rm c} \approx 0.2\,$cm/s exists in the sample, while the Landau critical velocity $v_{\rm cL} = 0$ in the polar phase.
} \label{SQVsVelocity}
\end{figure}

The measured magnon BEC relaxation rate $\tau_M^{-1}$ as a function of  $\Omega$, when a change of rotation velocity is started from the vortex-free state at $\Omega=0$, is presented in Fig.~\ref{SQVsVelocity}. It shows that there is a clear velocity, above which the relaxation rapidly grows. We interpret this point as a critical velocity for vortex formation. These vortices remain in the sample even when $\Omega$ is returned to zero, as expected if the vortices are pinned on the nafen strands. After the $\Omega$ cycle is finished, we identify the type of the formed vortices by measuring the cw NMR spectrum of the sample, Fig.~\ref{spectra}a. Since we find no satellite peak characteristic to HQVs, we infer that vortices formed during the $\Omega$ cycle are SQVs. In our experimental conditions the critical velocity for SQV formation is thus lower than that for forming HQVs. This is the case although the energy of an HQV pair is smaller than that of a single SQV in zero magnetic field (which is applied when $\Omega$ changes). A similar situation is observed in $^3$He-A, where the critical velocity for the double-quantum vortex skyrmions \cite{dqv} is lower than that of SQVs, while the energy preference is the opposite \cite{dqvsqv}.

In our cubic container the flow is non-uniform and at $\Omega= 1\,$rad/s, which is seen as a characteristic angular velocity for vortex formation in Fig.~\ref{SQVsVelocity}, the maximum flow velocity of 0.2\,cm/s is reached in the middle of each side wall of the square container cross-section. This value is somewhat lower than $v_{\rm c} \approx 1\,$cm/s observed for SQVs in bulk $^3$He-B, where it was also found to depend strongly on the surface conditions \cite{Bcritvel}. For confined samples control of the surface conditions, especially at the boundaries of the confining matrix, remains a challenge for the future.

Remarkably, we are also able to create HQVs in the superfluid state by changing the rotation velocity rapidly enough, see Fig.~\ref{spectra}c. To produce HQVs we vary the rotation velocity between $\Omega= + 2.25\,$rad/s and $-2.25\,$rad/s for several hours, in which case the SQV creation and annihilation is not able to compensate for changes in the rotation velocity fast enough and local flow velocity can exceed the critical velocity for HQV formation. While the amount of SQVs  created in this process increases magnon BEC relaxation beyond what can be measured, HQVs are also created, as seen from the appearance of the satellite in the cw NMR spectrum (magenta trace in Fig.~\ref{spectra}a). Therefore it is possible to set bounds for the HQV critical velocity for the conditions of the measurements: it exceeds 0.2\,cm/s but is below 1\,cm/s. 

Before the first vortices are formed, the superfluid is in a stable flow with non-zero velocity with respect to the walls of the container and to nafen strands, despite the fact that the Landau critical velocity is zero in the polar phase.

\section{Numerical simulations}

\begin{figure}[t]
\centerline{\includegraphics[width=0.9\linewidth]{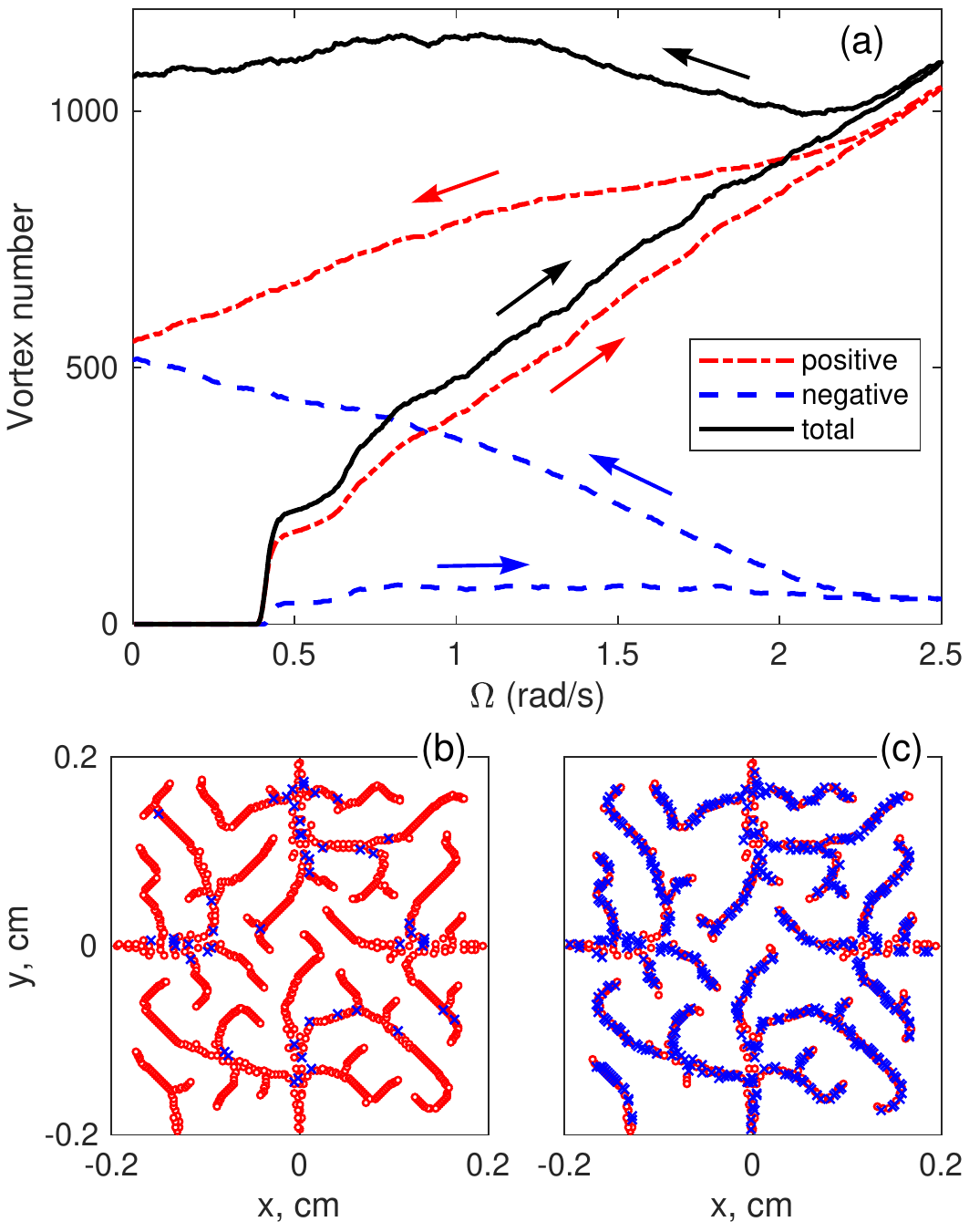}}
\caption{Simulation of the vortex formation with pinning: (a) Number of vortices (red dash dot line), antivortices (blue dash line) and the sum of the two populations (solid black line), as a function rotation velocity $\Omega$ which is changed from 0 to 2.5\,rad/s and back. (b) The configuration of vortices (red circles) and antivortices (blue crosses) at $\Omega=2.5\,$rad/s. (c) The final vortex configuration at $\Omega = 0$.
\label{simulation}
} 
\end{figure}

We can gain qualitative understanding of the process that creates vortices during $\Omega$ sweep in the superfluid state with a simple numerical model, Fig.~\ref{simulation}. In the model we consider a square sample container with $4 \times 4$~mm$^2$ cross section containing a grid of $201\times201$ pinning point and rotated about its center along the axis perpendicular to the cross section. The flow velocity is calculated as the sum of the potential flow in a box, and the contribution from vortices, which in the two-dimensional simulation are points. Vortices carry one quantum of circulation either along the rotation velocity or in the opposite direction (antivortices).

We start from the configuration with no vortices and zero rotation velocity $\Omega$. Then $\Omega$ is increased in small steps of $5\cdot 10^{-3}\,$rad/s. After each step we calculate the superfluid velocity in the rotating frame at each grid point. If the flow velocity magnitude exceeds $v_{\rm c,simul} = 0.1\,$cm/s (the imposed critical velocity in the simulation), a vortex or an antivortex is placed on the pinning site on the condition that the free energy of the system is lowered as a result. Vortices with positive and negative circulation at the same site annihilate each other. Once created, vortices are not allowed to move, which emulates the strong pinning by nafen observed experimentally \cite{HQVs_prl}. The flow field of the vortex is then added to the total flow.

In this model, the resulting vortex configuration forms a non-uniform pattern, Fig.~\ref{simulation}b,c which resembles that observed in superconducting thin-film systems where pinning is also strong \cite{dendrites,dendrites2,dendrites3}. The total number of vortices rapidly increases from zero when velocity of the potential flow exceeds $v_{\rm c,simul}$. These vortex avalanches emerge from the middle of each of the four container walls, where $v_{\rm c,simul}$ is first reached. This feature resembles the substantial jump in $\tau_M^{-1}$ at the critical velocity in the experiments. When the rotation velocity is later decreased, many vortices with positive circulation are annihilated by antivortices. However, the total number of vortices and antivortices remains relatively constant, which agrees with the observation that the magnon BEC relaxation rate never decreases with changes in the rotation velocity.

\section{Topology of the Bogoliubov Fermi Surface}

Detection of stable superflow in the polar phase is an indirect indication of the formation of a BFS, given by solution of $E({\bf p})=0$ in Eq.~\eqref{epsilon}. Fig.~\ref{polarfig}b  demonstrates the BFS for the superflow in the $(x,y)$ plane transverse to the $\unit m$ vector. Such BFS possesses quite remarkable features compared to the BFS expected to form in super-Landau flow in $^3$He-A or cuprate superconductors. In the case of $^3$He-A with the point nodes and in the available range of stable superflow velocities \cite{Ruutu1997}, the BFS is formed as tiny ellipsoidal pockets around the nodes. In case of cuprates, cylinders are formed around separate line nodes \cite{Volovik93,Tsuei2000}. Both cases are topologically trivial. In the polar phase, the BFS is formed by two (electron and hole) pockets, which extend across the whole momentum space even at the smallest velocities and touch each other at two points with non-trivial topology. Such Fermi surface resembles that in graphite, where the chain of touching electron and hole pockets is present \cite{Mikitik2006,Mikitik2008,Mikitik2014,HeikkilaVolovik2015a,Volovik2017}.

The non-trivial topology of the BFS in the polar phase is associated with the conical touching points at ${\bf p}= \pm p_{\rm F}\hat{\bf v}_{\rm s} \times \hat{\bf z}$. It is similar but not identical to that of the Weyl point in Weyl semimetals and in $^3$He-A. This follows from the Bogoliubov-de Gennes Hamiltonian:
\begin{equation}
H=  \tau_1 n_1({\bf p}) +n_2({\bf p})+  \tau_3 n_3({\bf p}) \,,
\label{Rashba2}
\end{equation}
where $\tau_1$ and $\tau_3$ are Pauli matrices in the particle-hole space. As distinct from the Weyl Hamiltonian, the matrix $\tau_2$ is missing and thus we call those points pseudo-Weyl points \cite{Volovik2017}. The components of the vector ${\bf n}({\bf p})$ are $n_1({\bf p})  = cp_z$,  $n_2({\bf p})=  {\bf p}\cdot{\bf v}_{\rm s}$ and $ n_3({\bf p}) =v_{\rm F}(p-p_{\rm F})$. The invariant, which is similar to that for the Weyl points, is:
\begin{equation}
N_3({\rm pseudo})=\frac{1}{8\pi}e_{ijk}\int_{S_2} dS^k ~\hat{\bf
n}\cdot \left(\frac{\partial \hat{\bf n}}{\partial {p_i}} \times \frac{\partial
\hat{\bf n}}{\partial {p_j}}\right) \,,
\label{InvariantPseudo}
\end{equation}
where $\hat{\bf n}= {\bf n} |{\bf n}|^{-1}$ is a unit vector and $S_2$ is the spherical surface around the touching point. Topological charges of the two pseudo-Weyl points are $N_3({\rm pseudo})=\pm 1$.

It is interesting that models of some superconducting states in heavy-fermion superconductors include closed node lines, like in the polar phase of $^3$He \cite{Joynt2002,Pfleiderer2009}. We thus suggest that topologically non-trivial BFS could be realized also in those systems, provided that pseudo-Weyl points turn out to be robust against impurities.

Let us now discuss predictions for observables resulting from the appearance of the non-thermal normal component in the polar phase. The BFS leads to a finite density of states (DoS):
\begin{equation}
N(0)= \int \frac{d^3 p}{(2\pi)^3} \delta(E({\bf p}))= N_{\rm F} \frac{\vs}{c} 
\,,
 \label{DOS}
\end{equation}
where $N_{\rm F}= p_{\rm F} m^*/\pi^2$ is the DoS in the normal $^3$He and $m^*=p_{\rm F}/v_{\rm F}$ is the effective mass. This results in a finite density of the normal component at $T=0$  and an additional heat capacity, which both are linear in $\vs$:
\begin{equation}
    \frac{\rhon (T=0)}{\rho} =  \frac{\vs}{c}  \frac {m^*}{m}\,\,,\, \,   \frac{C(T)}{C_{\rm F}(T)} =  \frac{\vs}{c}\,.
\label{rhon}
\end{equation}
Here $C_{\rm F}(T)$ is heat capacity of the normal liquid. For $v_{\rm s} \sim 0.2\,$cm/s the additional DoS is on the order of $0.05N_{\rm F}$, which, in principle, is detectable.
 
Additionally, the presence of superflow suppresses the gap amplitude. According to Muzikar and Rainer  \cite{MuzikarRainer1983}, the suppression of the gap at $T=0$ and $\vs \ll c$ is $(\Delta_{\rm P}(\vs)- \Delta_{\rm P}(0))/ \Delta_{\rm P}(0)=-\vs^3/3c^3$, and
at experimentally relevant temperatures
  \begin{equation}
\frac{\Delta (\vs,T)}{\Delta (0,0)}=\left[ 1-\alpha_1\frac{T^3}{T_{\rm c}^3}  
-\alpha_2  \frac{\vs^2}{c^2}  \frac{T}{T_{\rm c}}\right] \,, \,\vs/c \ll T/T_{\rm c}\ll1
\,,
 \label{GapSuppressionSauls2}
   \end{equation}
where the parameters $\alpha_1$, and $\alpha_2$ are of order of unity \cite{YipSauls1995}.
As a result, the spin-orbit interaction $F_D = g_D (\hat{\bf d}\cdot \hat{\bf m})^2$ is also suppressed with $\delta g_D/g_D=-2\vs^3/3c^3$ at $T=0$.
A well-established method to measure the strength of the spin-orbit interaction in superfluid $^3$He is through the shifts of the characteristic lines in the NMR spectra. The smallness of the expected effect (relative frequency shift $\sim 10^{-6}$ for the temperature and velocity reached in the present experiment) will, however, make this measurement challenging.

\section{Conclusions}

We have experimentally observed stable superflow in the polar phase of $^3$He at velocities exceeding the zero Landau critical velocity in the nodal line superfluid. The stability of superflow, provided by formation of the Bogoliubov Fermi surface and of the non-thermal normal component at super-Landau velocities, is limited by creation of the single-quantum vortices (at velocities of about 0.2\,cm/s) and of the half-quantum vortices (at velocities below 1\,cm/s). The next development will be to observe the contribution of the flow-induced quasiparticle states to thermodynamic quantities, e.g.\ those predicted by Eqs.~(\ref{rhon}) and (\ref{GapSuppressionSauls2}). We thus confirm that appearance of protected zero-energy states in topological superfluids and superconductors does not prevent existence of stable superflow in such systems. When the the original zero-energy states belong to a closed line node, we predict that the resulting BFS possesses non-trivial topology with the pseudo-Weyl points. It will be appealing to find a topologically non-trivial BFS existing even without the applied flow in systems with broken symmetries. It will also be interesting to elucidate further consequences of the symmetries, broken by the superflow in topological superfluids and superconductors, beyond formation of the BFS. One example here is provided by  the prediction of the spin-stripe phases \cite{Brauner2019,Volovik2020}.

In the case where BFS is formed by the flow at super-Landau velocities, like in the polar phase, the initial process of filling the negative energy states is also a fascinating problem for future research, as it proceeds via radiation of quasiparticles which has similarities to the Hawking radiation from the black-hole horizon \cite{VolovikHawking,Kedem2020,Liang2019}. Finally we note that recent successes in stabilization of uniform ultracold quantum gases \cite{UniformGas} open possibilities to study the evolution of the super-Landau superflow \cite{Yamamura2015} in the BEC-BCS crossover, where the spectrum of Bogoliubov excitations in the stationary system changes drastically.

\begin{acknowledgments}
We thank V.V. Dmitriev for instructive discussions and providing the nafen sample.
This work has been supported by the European
Research Council (ERC) under the European Union's
Horizon 2020 research and innovation programme (Grant
Agreement No. 694248). The experimental work was
carried out in the Low Temperature Laboratory, which
is part of the OtaNano research infrastructure of Aalto
University. S. Autti acknowledges financial support from
the Jenny and Antti Wihuri foundation.
\end{acknowledgments}

S.A. and J.M. contributed equally to this work.

\def\He#1{$^{#1}$He}

\end{document}